\documentclass[%
 reprint,
superscriptaddress,
showpacs,preprintnumbers,
 amsmath,amssymb,
 aps,
pra,
]{revtex4-1}

\usepackage{graphicx}
\usepackage{dcolumn}
\usepackage{bm}

\usepackage{mathptmx}
\usepackage[colorlinks,linkcolor=blue, urlcolor=blue, anchorcolor=blue, citecolor=blue]{hyperref}
\begin{document}

\preprint{APS/123-QED}

\title{Probing vectorial near field of light: imaging theory and design principles of nanoprobes}

\author{Lin Sun} 
\email{sunlin13@mails.tsinghua.edu.cn}
\affiliation{%
State Key Laboratory of Precision Measurement Technology and Instruments, \\
Department of Precision Instrument, Tsinghua University, Beijing 100084, China 
}%

\author{Benfeng Bai}%
 \email{baibenfeng@tsinghua.edu.cn}
\affiliation{%
State Key Laboratory of Precision Measurement Technology and Instruments, \\
Department of Precision Instrument, Tsinghua University, Beijing 100084, China 
}%
\affiliation{%
Tsinghua-Foxconn Nanotechnology Research Center, Tsinghua University, Beijing 100084, China 
}%

\author{Jia Wang}
\affiliation{%
State Key Laboratory of Precision Measurement Technology and Instruments, \\
Department of Precision Instrument, Tsinghua University, Beijing 100084, China 
}%

\date{\today}

\begin{abstract}
Near-field microscopy is widely used for characterizing electromagnetic fields at nanoscale, where nanoprobes afford the opportunity to extract subwavelength optical quantities, including the amplitude, phase, polarization and chirality. However, owing to the complexity of various nanoprobes, a general and intuitive theory is highly needed to assess the vectorial field response of the nanoprobes and interpret the mechanism of the probe-field interaction. Here, we develop a general imaging theory based on the reciprocity of electromagnetism and multipole expansion analysis. The proposed theory closely resembles the multipolar Hamiltonian for light-matter interaction energy, revealing the coupling mechanism of the probe-field interaction. Based on this theory, we introduce a new paradigm for the design of functional nanoprobes by analyzing the reciprocal dipole moments, and establish effective design principles for the imaging of vectorial near fields. Moreover, we numerically analyze the responses of two typical probes, which can quantitatively reproduce and well explain the experimental results of previously reported measurements of optical magnetism and transverse spin angular momentum. Our work provides a powerful tool for the design and analysis of new functional probes that may enable the probing of various physical quantities of the vectorial near field.
\end{abstract}

\maketitle


\section{\label{sec:level1}Introduction}

In the past decades, scanning near-field optical microscopy (SNOM) has been widely applied to study the electromagnetic properties of nanostructures with deep subwavelength resolution \cite{RN167}. Besides the abilities of surface topographic imaging and near-field spectroscopy imaging, the most compelling feature of SNOM is the privilege to detect the complex vectorial electromagnetic field. Thus far, several novel functional probes and related techniques have been proposed to facilitate the near-field imaging of physical quantities of light, such as the imaging of electric vector by probes with phase-resolved techniques \cite{RN322,RN320,RN321} or by probes adhered with nanoparticles \cite{RN323}. Moreover, the vectorial imaging of weak magnetic field at optical frequencies has been achieved by split-ring probes \cite{RN309} for the out-of-plane magnetic component, by hollow pyramid probes \cite{RN313,RN311} or Bethe-hole probes \cite{RN310,RN312} for the in-plane magnetic components, and even by probes with rare-earth ions for magnetic transitions \cite{RN175}. Recently, the transverse spin angular momentum (SAM) of evanescent field has been detected by scanning nanoparticles \cite{RN87,RN303}. Therefore, SNOM has become a powerful tool and provides a direct route to probe, analyze and interpret the underlying physics in nano-optics.

Along with the development of SNOM, theories in near-field optics have been established for interpreting the imaging process by SNOM. In the early stage, the probe was regarded as an electric dipole and the detected signal was regarded proportional to the Poynting vector of local fields \cite{RN333}. Though working well for some applications, this model does not consider the influence of the material and geometry of the probe, and even neglects the response to the magnetic field of light. Some numerical methods were also developed \cite{RN334}, for both the collection mode and the illumination mode of SNOM \cite{RN275}. Remarkably, the reciprocity of electromagnetism was first explicitly considered by Greffet and Carminati \cite{RN330} for discussing the equivalence of the collection mode and the illumination mode. Since then, imaging theories based on reciprocity have been established, such as Bardeen's formulas for SNOM \cite{RN329,RN344} and the thermal emission spectrum for electromagnetic local density of states (LDOS) \cite{RN328}. Most importantly, a virtual reciprocal scenario was introduced to analyze the actual experimental scenario with SNOM by reciprocity \cite{RN197}, which reveals the vectorial response of the probe. The reciprocity theory with these two scenarios provides a general framework for both aperture-type SNOM and apertureless SNOM and significantly extends the range of applications. In the weak-coupling regime for tip-sample interaction, the near-field imaging by SNOM can be approximated as a convolution process according to the reciprocity theory, which has been corroborated by many near-field measurements \cite{RN327,RN202,RN199}. Recently, some variations of the reciprocity theory have been proposed, such as a convolution model of intensity imaging for the aperture-type SNOM \cite{RN153} and a vectorial model of complex amplitude imaging for the apertureless SNOM \cite{RN201}. Notably, the convolution model originating from the reciprocity theory is not sufficiently valid for applications in strong-coupling regime, where the probe has a remarkable influence on local electromagnetic fields \cite{RN298} or spectral features \cite{RN299}. In such situations, the Green dyadics is a sound theoretic tool to analyze the interaction \cite{RN198} and even to calculate the field distributions \cite{RN300}.

The nanoprobe is a vital element in SNOM, acting as an optical antenna \cite{RN118} that converts the propagating waves into evanescent waves and vice versa. Therefore, an important question in imaging of vectorial near field by SNOM is whether the probe is electric-sensitive or magnetic-sensitive. The reciprocity theory establishes a relation between the probe and the near field, and seemingly indicates that the probe simultaneously responses to both the electric field and magnetic field \cite{RN199}. However, this implicit relation provides less quantitative information to guide the design of novel functional probes or to analyze the vectorial field imaging without priori knowledge. Thus, the relevant questions addressed in this paper are: whether an imaging theory exists for determining whether the dominant contribution of the detected signal is from the electric field or magnetic field of light, and whether a solid figure of merit (FOM) can be established to estimate the vectorial responses of the probe to electromagnetic fields. Traditionally, the multipole expansion of optical antennas is an effective tool to study the far-field radiation of antennas \cite{RN173,RN331}. Recently, this technique was also used to analyze the inherent polarizability \cite{RN297} and near-field coupling effect \cite{RN295} of optical antennas. Inspired by Ref. [\citenum{RN332}], where near-field imaging by the aperture-type SNOM was regarded as waveguide coupling with dipole moments, we introduce multipole expansion of the probe to reciprocity of electromagnetism to address the above questions.

In this work, we establish a general imaging theory of nanoprobes based on reciprocity of electromagnetism and multipole expansion analysis. Strikingly, this theory is in close analogy with the multipolar Hamiltonian \cite{RN175,RN173,RN290} for light-matter interaction energy, explicitly indicating that the SNOM signal is proportional to the exact coupling power between the probe and the near fields. For facilitating the applications, the coupled moment model (CMM) is directly derived from this theory. Based on the CMM, we propose a FOM that represents the ratio between the magnetic and electric dipole moments in the virtual reciprocal scenario. Referring to this FOM and the reciprocal dipole moments, one can easily assess the vectorial response of the probe, which leads to design principles for novel functional probes. By comparing with prior experiments \cite{RN309,RN87,RN303}, two imaging processes of vectorial near field are investigated numerically. One is to probe the optical magnetism at near-infrared spectrum with a split-ring probe, and the other is to detect the transverse SAM of light in the visible range with a nanoparticle probe. The consistencies between calculations and experiments demonstrate the effectiveness of the proposed models and principles, which could be promising tools for understanding and analyzing vectorial near-field detections.

\section{\label{sec:part2}imaging theory based on\protect\\ reciprocity and multipole expansion}  
Throughout this paper, we consider monochromatic electromagnetic fields and use SI units. All the materials are non-optically active and non-magnetic (i.e., $\mu_r=1$). Without loss of generality, we apply the complex dielectric tensor to represent the electromagnetic properties of materials. Moreover, Einstein summation convention and Levi-Civita symbol are adopted for the representation of the vectors with spatial index for simplicity.

\subsection{Imaging theory with Hamiltonian and reciprocity}
Hamiltonian quantifies the interaction between particles (or system) and the external electromagnetic field \cite{RN290}. For system smaller than the wavelength, Hamiltonian is conveniently described as the multipolar formula $\hat{H}_{\rm{int}} = -\bm{p}\cdot\bm{E} - \bm{m}\cdot\bm{B} - \bm{Q}:\bm{\nabla}\bm{E} + \cdots$ in the appropriate gauge \cite{RN173,RN359}. Analogously, the probe-field interaction and the imaging process in near-field microscopy or SNOM can also be associated with multipolar Hamiltonian in the framework of reciprocity theory \cite{RN197}.

\begin{figure}[h]
\centering\includegraphics{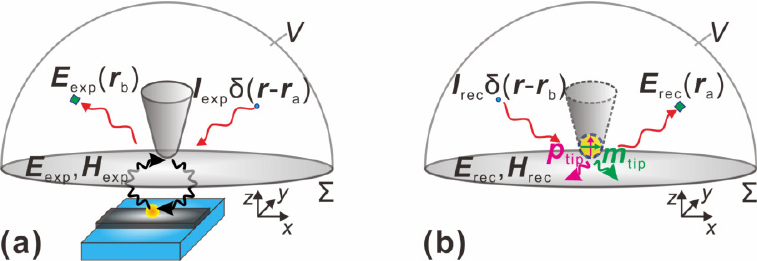}
\caption{\label{fig:1} Schematic of reciprocity in near-field optics. (a) Experimental scenario: a source $\bm{I}_{\rm{exp}}$ located at $\bm{r}_{\rm{a}}$ illuminates the tip-sample system, and a detector at $\bm{r}_{\rm{b}}$ obtains the far-field signal $\bm{E}_{\rm{exp}}$. Near fields $\{\bm{E}_{\rm{exp}},\bm{H}_{\rm{exp}}\}$ in an evaluation plane $\Sigma$ are determined by the probe and sample simultaneously. (b) Reciprocal scenario: a source  $\bm{I}_{\rm{rec}}$ at $\bm{r}_{\rm{b}}$ emits radiation, and a detector at $\bm{r}_{\rm{a}}$ gets the far-field signal $\bm{E}_{\rm{rec}}$. Near fields $\{\bm{E}_{\rm{rec}},\bm{H}_{\rm{rec}}\}$ in $\Sigma$ are solely determined by the probe, which could be approximated by a series of multipolar moments (e.g., the electric dipole moment $\bm{p}_{\rm{tip}}$, the magnetic dipole moment $\bm{m}_{\rm{tip}}$, etc.).}  
\end{figure}

Reciprocity describes the relation between two sets of electromagnetic fields under two reciprocal scenarios (cf. Fig. \ref{fig:1}\textcolor{blue}). The experimental scenario with subscript $exp$ is an actual SNOM setup for probing near-fields $\bm{E}_{\rm{exp}}(\bm{R})$ and $\bm{H}_{\rm{exp}}(\bm{R})$ in the near-field plane $\Sigma$, while the reciprocal scenario with subscript $rec$ corresponds an imaginary setup for providing near-fields $\bm{E}_{\rm{rec}}(\bm{R})$ and $\bm{H}_{\rm{rec}}(\bm{R})$ as a priori knowledge of the nanoprobe. In both scenarios, the material distribution is identical in the volume $V$ that contains the nanoprobe except for interchanging the location of the source and the detector.

Supposing that the material in volume $V$ is non-optically active and non-magnetic, and the plane $\Sigma$ is entirely in vacuum, the reciprocity can be expressed as \cite{RN198} (Appendix \ref{sec:appendix_a})
\begin{equation}
\begin{split}
\bm{E}_{\rm{exp}}\cdot\bm{I}_{\rm{rec}}|_{\bm{r}_{b}}& - \bm{E}_{\rm{rec}}\cdot\bm{I}_{\rm{exp}}|_{\bm{r}_{a}}  \\
&= \frac{1}{i\omega\mu_0}\int_{\Sigma}dS(E_{2,j}\partial_z{E_{1,j}} - E_{1,j}\partial_z{E_{2,j}}) .   \label{equ_1}
\end{split}
\end{equation}
Here, the subscripts $rec$ and $exp$ are replaced by 1 and 2 respectively for simplicity. Equation (\ref{equ_1}) is akin to Bardeen's formula for scanning tunnel microscope (STM) \cite{RN329,RN344,RN328}. The left terms in Eq. (\ref{equ_1}) are commonly referred to as the mutual impedance \cite{RN199}, where the first and second term correspond to the far-field detected signal and the background noise, respectively. Mostly,  the noise term $\bm{E}_{\rm{rec}}\cdot \bm{I}_{\rm{exp}}|_{\bm{r}_{\rm{a}}}$ can be effectively suppressed by three strategies: First, the external source can be out of the volume $V$ (e.g., using SNOM in transmission mode or setting excitation by waveguides or prism coupling); Second, the cross-polarized technique can be applied [i.e., the polarization of $\bm{I}_{\rm{exp}}(\bm{r}_a)$ is orthogonal to that of $\bm{E}_{\rm{rec}}(\bm{r}_a)$]; Third, modulation and demodulation process can be used by tapping the cantilever in apertureless SNOM \cite{RN201}. Thus, we neglect this noise term, and use $\cal{M}_{\rm{pol}}$$(\bm{r})$ to replace the signal term $\bm{E}_{\rm{exp}}\cdot\bm{I}_{\rm{rec}}|_{\bm{r}}$, where the subscript $pol$ represents the polarization of $\bm{I}_{\rm{rec}}$. 

For emphasizing the contribution from the nanoprobe explicitly, applying the scattering theory to Eq. (\ref{equ_1}), one obtains reciprocity in a new fashion (Appendix \ref{sec:appendix_a})
\begin{equation}
\mathcal{M}_{\rm{pol}}(\bm{r}_{\rm{tip}}) = \int_{V'}d^3r'\bm{S}^{(+)}_\textit{E}(\bm{r}')\cdot\bm{J}_{\rm{tip,P}}(\bm{r}'),   \label{equ_2}
\end{equation}
where $V'$ is a subdomain in $V$ that contains the whole nanoprobe. Here, $\bm{J}_{\rm{tip,P}}$ is the polarization current density of the probe in the reciprocal scenario and satisfies $\bm{J} = \dot{\bm{P}} = -\textit{i}\omega\varepsilon_0\bm{\chi}\cdot\bm{E} = -\textit{i}\omega\varepsilon_0(\tilde{\bm{\varepsilon}}_r - \bm{I})\cdot\bm{E}$ where $\bm{\chi}$, $\tilde{\bm{\varepsilon}}_r$ and $\bm{I}$ represent the electric susceptibility tensor, the complex relative dielectric tensor and the unit dyadics. $\bm{S}_E^{(+)}(\bm{r}')$ in Eq. (\ref{equ_2}) is the upward component of a $synthetic\;field$ $\bm{S}_{\Psi}(\bm{r}')$ with $\Psi = \{E, H\}$, which satisfies
\begin{equation}
\bm{S}_\Psi(\bm{r}) = \sum_{\xi = \pm 1}\frac{1}{2\pi}\int{d^2K\bm{S}_\Psi^{\xi}(\bm{K},z_0)\exp[i\bm{K}\cdot\bm{R}+i\xi\gamma(z - z_0)]},  \label{equ_3}
\end{equation}
where $z_0$ is the $z$-coordinate of the evaluation plane $\Sigma$ and the angular spectra satisfy $\bm{S}_{\Psi}^{\xi}(\bm{K},z_0) = \bm{\Psi}^{\xi}(\bm{K},z_0)$ with superscript $\xi = +1$ and $-1$ corresponding to the upward and downward components, respectively. Mathematically, the upward and downward components can be separated by two consecutive angular spectra of $\bm{S}_{\Psi}$ with distance $\Delta z$  
\begin{equation}
\bm{S}_\Psi^{\xi}(\bm{K},z) = \xi\frac{\bm{S}_\Psi(\bm{K},z + \Delta z) - \bm{S}_\Psi(\bm{K},z)\exp(-i\xi\gamma\Delta z)}{2\sinh(i\gamma\Delta z)}.        \label{equ_4}
\end{equation}
Here, $synthetic$ means that the field $\bm{S}_{\Psi}(\bm{r})$ is a calculated field based on $\bm{S}_{\Psi}(\bm{R},z_0)$ and Eq. (\ref{equ_3}), rather than the actual field $\bm{E}(\bm{r})$ or $\bm{H}(\bm{r})$ existing in the tip-sample system, especially for regions out of the vacuum between the tip and the sample.

For estimating the electromagnetic response of the nanoprobe more explicitly (i.e., whether the probe is electric-sensitive or magnetic-sensitive), the multipole expansion analysis of $\bm{J}_{\rm{tip,P}}$ is introduced to reciprocity theory. For a small volume probe (i.e., satisfying $|\bm{k}| |\bm{r}| \leqslant 1$ where $\bm{r}$ is inside the tip volume), one can cast Eq. (\ref{equ_2}) in a general form after some algebraic manipulations (Appendix \ref{sec:appendix_b})
\begin{equation}
\begin{split}
\mathcal{M}_{\rm{pol}}(\bm{r}_{\rm{tip}}) = &- i\omega\bm{p}_{\rm{tip}}\cdot\bm{S}_{\textit{E}}^{(+)}(\bm{r}_{\rm{tip}}) + i\omega\bm{m}_{\rm{tip}}\cdot\bm{S}_{\textit{B}}^{(+)}(\bm{r}_{\rm{tip}}) \\
 &- \frac{1}{6}i\omega\bm{Q}_{\rm{tip}} : \bm{\nabla}\bm{S}_{\textit{E}}^{(+)}(\bm{r}_{\rm{tip}}) + \cdots ,  \label{equ_5}
\end{split}
\end{equation} 
where $\bm{S}_{B} = \mu_0 \bm{S}_{H}$. Here, $\bm{p}_{\rm{tip}}$, $\bm{m}_{\rm{tip}}$ and $\bm{Q}_{\rm{tip}}$ are the electric dipole moment, the magnetic dipole moment and the electric quadrupole moment tensor of the tip in the reciprocal scenario, denoted as the $reciprocal\;multipole\;moments$ and defined as  $\bm{p} = \int_{V'}d^3r'\rho(\bm{r}')\bm{r}'$, $\bm{m} =(1/2)\int_{V'}d^3r'\bm{r}'\times\bm{J}(\bm{r}')$ and $\bm{Q} = \int_{V'}d^3r'\{ (3\bm{r}'\bm{r}' - r'^2 \bm{I})\rho(\bm{r}') \}$, respectively, associated with the charge $\rho (\bm{r}')$ and induced current density $\bm{J} (\bm{r}')$ of the tip in the reciprocal scenario \cite{RN335}.

Equation (\ref{equ_5}) is the exact formulation of our proposed imaging theory for SNOM and uses reciprocal multipole moments of the probe to replace the reciprocal field, which can explicitly determine the contributions from the near-field electric field $\bm{S}_{E}$, the magnetic field $\bm{S}_B$, the gradient of electric field $\bm{\nabla}\bm{S}_E$, etc. Remarkably, equation (\ref{equ_5}) closely resembles the multipolar expansion of Hamiltonian for light-matter interaction energy \cite{RN175,RN290,RN173}, interpreting the coupling mechanism between the nanoprobe and the near field to be measured. Supposing an interaction Hamiltonian in the form $\hat{H}_{\text{int}} = \bm{p}\cdot\bm{E} - \bm{m}\cdot\bm{B} + (1/6)\bm{Q}:\bm{\nabla}\bm{E} + \cdots$, equation (\ref{equ_5}) is exactly the time derivative of the interaction Hamiltonian for time-harmonic fields
\begin{equation}
\cal{M}_{\rm{pol}} = \frac{\rm{1}}{\rm{2}} \dot{\hat{\textit{H}}}_{\rm{int}}, \label{equ_6}
\end{equation}
which reveals that the far-field detected signal $\cal{M}_{\rm{pol}}$ obtained by SNOM unambiguously equals to half of the near-field coupling power $\partial{\hat{\textit{H}}}_{\rm{int}} / \partial t$ between the tip and the sample. Equations (\ref{equ_5}) and (\ref{equ_6}) describe the near-field imaging as a power coupling process between the tip and the sample where the vectorial response is associated with the reciprocal multipole moments, and underscores the fundamental physical insights in near-field microscopy.

\subsection{Coupled moment model and vectorial FOM}
Equations (\ref{equ_5}) and (\ref{equ_6}) provide a general model for SNOM that contains both the information of the tip-sample interaction from the experimental near-fields (i.e., $\bm{S}_{E}$, $\bm{S}_{B}$, $\bm{\nabla}\bm{S}_{E}$, etc.) and that of the probe's vectorial response from the reciprocal multipole moments (i.e., $\bm{p}_{\rm{tip}}$, $\bm{m}_{\rm{tip}}$, $\bm{Q}_{\rm{tip}}$, etc.). For the influence of the higher-order tip-sample interaction, an intriguing and elaborate analysis has been conducted in Ref. [\citenum{RN198}]. In the following, we limit our concentration on the response of the nanoprobe to the vectorial near-fields. Thus, the higher-order tip-sample interaction is neglected, and the experimental near-fields can be approximated by the unperturbed sample near-fields as $\bm{S}_E^{(+)}(\bm{r}_{\rm{tip}}) \approx \bm{E}_{\rm{sample}}^{(+)}(\bm{r}_{\rm{tip}})$ and $\bm{S}_B^{(+)}(\bm{r}_{\rm{tip}}) \approx \mu_0 \bm{H}_{\rm{sample}}^{(+)}(\bm{r}_{\rm{tip}})$, according to the perturbation assumption \cite{RN167} commonly used in SNOM. Then, one can cast Eq. (\ref{equ_5}) in the form
\begin{equation}
\begin{split}
\mathcal{M}_{\rm{pol}}(\bm{r}_{\rm{tip}}) \approx &- i\omega\bm{p}_{\rm{tip}}\cdot\bm{E}_{\rm{sample}}^{(+)}(\bm{r}_{\rm{tip}}) + i\omega\bm{m}_{\rm{tip}}\cdot\bm{B}_{\rm{sample}}^{(+)}(\bm{r}_{\rm{tip}})  \\ 
&- \frac{1}{6}i\omega\bm{Q}_{\rm{tip}} : \bm{\nabla}\bm{E}_{\rm{sample}}^{(+)}(\bm{r}_{\rm{tip}}) + \cdots ,  \label{equ_7}
\end{split}
\end{equation} 
which is the coupled moment model (CMM) of reciprocity with perturbation assumption. Considering that the reciprocal dipole moments of nanoprobes are commoly dominant for SNOM, equation (\ref{equ_7}) can be further truncated and simplified as
\begin{equation}  \label{equ_8}
\mathcal{M}_{\rm{pol}}(\bm{r}_{\rm{tip}}) \approx -i\omega\bm{p}_{\rm{tip}}\cdot\bm{E}_{\rm{sample}}^{(+)}(\bm{r}_{\rm{tip}}) +i\omega\frac{\bm{m}_{\rm{tip}}}{c}\cdot Z_0\bm{H}_{\rm{sample}}^{(+)}(\bm{r}_{\rm{tip}}) ,
\end{equation}
where $Z_{\rm{0}}$ and $c$ are the impedance and the speed of light in vacuum, respectively. This simplified CMM is akin to the point-like tip model in Ref. [\citenum{RN198}] that only considers the near-field response of the probe to the electric field. Noteworthily, the proposesd model also contains a term corresponding to the magnetic field.  This difference originates from the volume of the tip. In the point-like tip model, the volume of the tip is so small (i.e., $|\bm{k}||\bm{r}| \ll 1$) that only the electric dipole moment exists effectively. However, the volume in CMM can be large enough (i.e., $|\bm{k}||\bm{r}| \sim 1$) to support an effective magnetic dipole moment. Thus, the CMM generalizes and extends the point-like tip model and interprets the mechanism for probing of the optical magnetism.

The CMM provides a theoretical tool to quantitatively analyze the contribution from either the electric field or the magnetic field. According to Eq. (\ref{equ_8}), the detected signal relies on both the near- field $\bm{E}_{\rm{sample}}$ or $\bm{H}_{\rm{sample}}$ of the sample and the corresponding reciprocal dipole moment $\bm{p}_{\rm{tip}}$ or $\bm{m}_{\rm{tip}}$ of the nanoprobe. For estimating the vectorial response of the tip, we take the electromagnetic field of a plane wave as the sample field (i.e., $\bm{E}_{\rm{sample}} = Z_0 \bm{H}_{\rm{sample}}$) without loss of generality. Then, we introduce a quantity $BE$ (named BE value) to describe the ratio between the magnetic coupling power and the electric counterpart, which is defined as
\begin{equation}   \label{equ_9}
BE(\omega) = 10 \log _{10} \bigg{\{}\frac{1}{c}\frac{|\bm{m}_{\rm{tip}}(\omega)|}{|\bm{p}_{\rm{tip}}(\omega)|}\bigg{\}}
\end{equation}
whose unit is dB. By the definition of Eq. (\ref{equ_9}), the nanoprobe is magnetic-sensitive when $BE(\omega)$ is positive, while it is electric-sensitive when $BE(\omega)$ is negative. Hence, $BE(\omega)$ is an effective FOM to predict or determine the vectorial response of the probe.

\subsection{Reciprocal dipole moments and polarizability tensor}
The polarizability tensor is a key quantity that represents the vectorial response of an optical antenna or a nanoprobe. Though containing all the vectorial information of the nanoprobe, the polarizability tensor is fully derived from the induced dipole moments by six calculations with orthogonal plane wave excitations \cite{RN358}, not to mention that the superpolarizability \cite{RN297} that also associate the electric quadrupole moment with the gradient of electric fields may be more complex to calculate. On the other hand, the dipole moments are inherently associated with the polarizability tensor \cite{RN290,RN296}. However, dipole moments of the probe in the experimental scenario are spatially dependent because these dipole moments are directly induced by the nanoscale near fields, severely hindering the applications of dipole moments. Fortunately, the reciprocal dipole moments are solely excited by a plane wave whose excitation conditions (e.g., the polarization state and the incident direction) are determined by the SNOM system in an actual experimental scenario. Therefore, the reciprocal dipole moments are spatially independent. To conclude, both the polarizability tensor and the reciprocal dipole moments are intrinsic quantities to represent the vectorial response of the probe. The polarizability tensor has complete vectorial information of the probe, but with a relatively complicated process to calculate. In contrast, the reciprocal dipole moments not only include enough vectorial information of the probe by one single calculation, but also consider the impact of the actual SNOM configuration, indicating CMM is an effective tool for interpreting and analyzing the vectorial imaging of optical near fields by SNOM.

\section{\label{sec:part3}design principles of nanoprobes and applications for probing vectorial near field}
\begin{figure}[h]
\centering\includegraphics{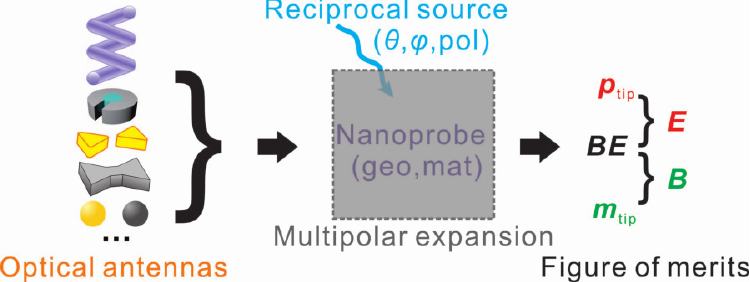}
\caption{\label{fig:2} Schematic of the design paradigm for the nanoprobes. Depending on the basic vectorial features of the optical antennas, a nanoprobe with preliminary parameters of material $mat$ and geometry $geo$ is selected. According to the SNOM scheme in the experimental scenario, the incident angle $(\theta,\;\phi)$ and polarization state $pol$ of the reciprocal source are set and this source excites the volume of the nanoprobe (i.e., the gray box). The reciprocal dipole moments $\bm{p},\;\bm{m}$ and the FOM $BE(\omega)$ are calculated by introducing multipole expansion to the nanoprobe. These FOMs can be regarded as feedbacks to optimize the design of the nanoprobes.}
\end{figure}

Based on the proposed theory and model, a design paradigm of nanoprobes for detecting the vectorial near field is proposed (cf. Fig. \ref{fig:2}). First, functional nanostructures or optical antennas [e.g., helix, split-ring resonator(SRR), bow-tie structure, diabolo, nanoparticles, etc.] are chosen with preliminary materials and geometric parameters. Then, multipole expansion analysis of the selected nanostructure is conducted by numerical simulations, where the excitation conditions [i.e., polarization state and incident direction] of the reciprocal source is determined by the SNOM configuration in the experimental scenario. Next, according to the following design principles, the reciprocal dipole moments ($\bm{p}_{\rm{tip}},\;\bm{m}_{\rm{tip}}$) and the derived FOM $BE(\omega)$ can be regarded as feedback parameters to optimize the nanoprobe and the SNOM scheme. Finally, the performances of the nanoprobe and SNOM scheme can be simply tested by CMM in Eq. (\ref{equ_8}), or even elaborately verified by rigorous reciprocity (RR) in Eqs. (\ref{equ_1}) and (\ref{equ_2}) where $rigorous$ means that no approximation is adopted to reciprocity of electromagetism.

The design principles of nanoprobes can be readily obtained based on the coupling expression Eq. (\ref{equ_7}) or Eq. (\ref{equ_8}). On one hand, a larger $\bm{p}_{\textup{tip}}$ or $\bm{m}_{\textup{tip}}$ makes the probe more electric-sensitive or magnetic-sensitive, respectively. Moreover, if one component of $\bm{p}_{\textup{tip}}$ or $\bm{m}_{\textup{tip}}$ is much larger than the others, the probe ought to measure the corresponding component of the electric field or magnetic field. Alternatively, if the $\bm{p}_{\textup{tip}}$ or $\bm{m}_{\textup{tip}}$ has nearly the same components, the probe can detect the electric or magnetic vector of the near fields. On the other hand, probes usually have responses to both the electric and magnetic fields. Thus, the electric and magnetic fields can hardly be extracted \cite{RN199,RN6} without $a\;priori$ knowledge of the sample fields, e.g., the symmetry of the waveguide modes \cite{RN309,RN199} and the interleaving distribution of tightly focused cylindrical-vector beams \cite{RN175,RN312}. Towards achieving vectorial imaging of arbitrary near fields with little $a\;priori$ knowledge, $BE(\omega)$ should be assessed to estimate the electromagnetic response of the probe. Specifically, the more positive the $BE(\omega)$ of the probe is, the higher signal to noise ratio (SNR) can be achieved for probing the weak magnetic field. In contrast, the more negative the $BE(\omega)$ is, the higher SNR can be achieved for imaging the electric field.

In the following subsections, two applications are demonstrated by using the two above-mentioned principles. The first one is to probe verticle component of the magnetic field (i.e., $H_z$) of a ridge waveguide at near-infrared wavelengths by a split-ring probe that has a large reciprocal magnetic dipole moment $m_{\textup{tip,}\;z}$ \cite{RN309}. The second one is to detect the transverse SAM of tightly focused cylindrical-vector beams at visible spectrum by a nanoparticle probe where the proposed FOM $BE(\omega)$ matters \cite{RN87,RN303}.

\subsection{Optical magnetism detection by a split-ring probe}
We start with an analysis of optical magnetism probing at a wavelength of $1.55\;\mu \rm{m}$ by a split-ring probe reported in Ref. [\citenum{RN309}]. Due to the evanescent feature of the fundamental TE mode in a ridge waveguide (RW), it is valid enough to apply a SRR thicker than the decay length to replace the split-ring probe. We use similar dimensions and parameters as in Ref. [\citenum{RN309}]. The SRR takes 250 nm radius for the outer aluminum layer, 100 nm radius for the inner glass core, 40 nm for the split width etched through the aluminum layer, 150 nm thickness for the whole SRR, and $\tilde{n} = 1.44 - i16.0$ for refractive index of aluminum \cite{RN360}. The gap between SRR and RW is 20 nm. The RW takes $2\;\mu \rm{m}$ thickness for the glass substrate, 280 nm thickness for the silicon nitride layer, 20 nm thickness for the silicon nitride ridge with $2\;\mu \rm{m}$ width, $\tilde{n} = 1.9$ for the refractive index of silicon nitride. The effective index for the fundamental TE mode approximately equals 1.55. 

\begin{figure*}
\centering\includegraphics{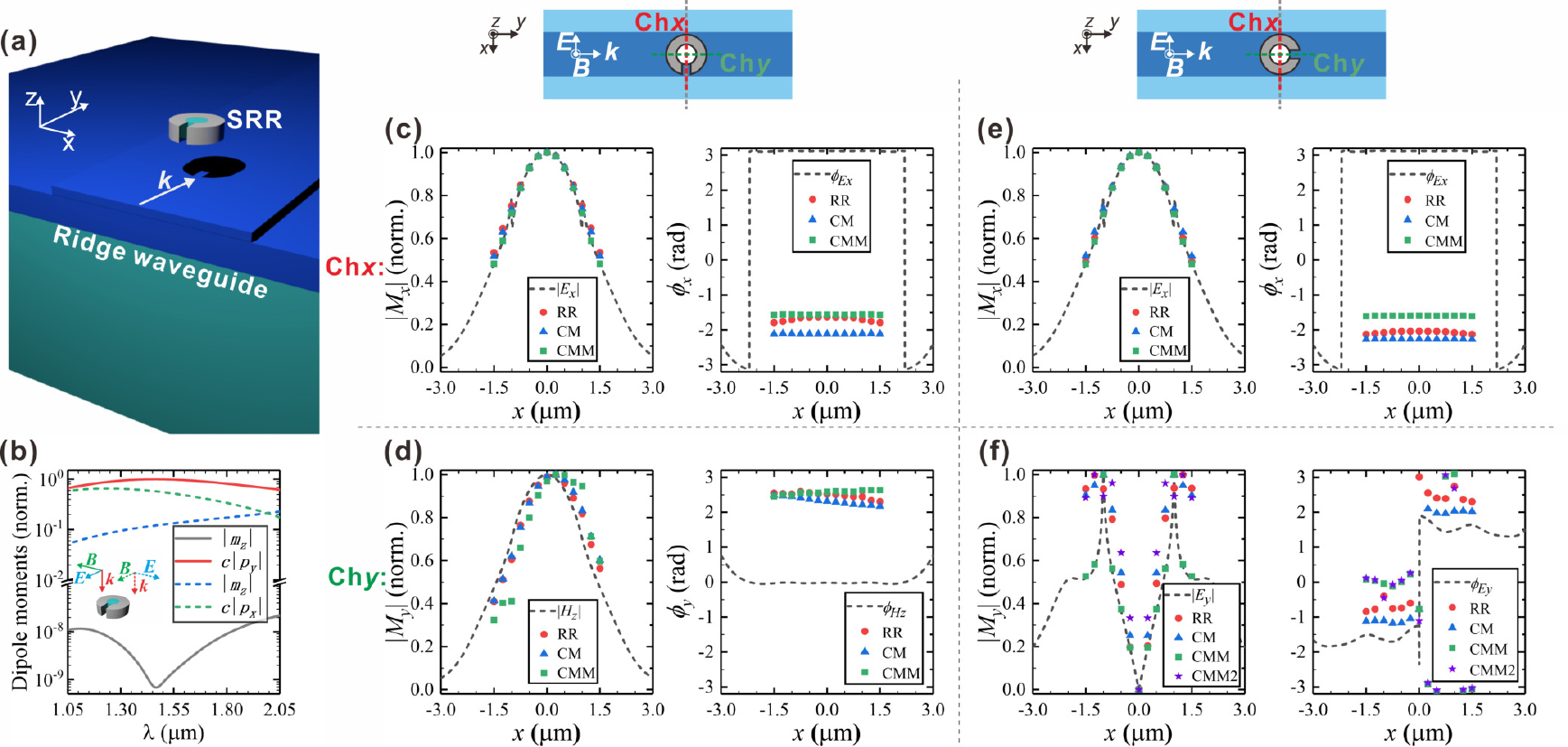}
\caption{\label{fig:3} Simulations of the optical magnetism detection with the split-ring probe. (a) Schematic of the tip-sample system. The split-ring probe is approximated by a split-ring resonator (SRR). (b) Distributions of the reciprocal dipole moments of the SRR. The solid and dashed curves are for circumstances where the electric field $\bm{E}$ is parallel and perpendicular to the facets of the split, respectively. (c,d) and (e,f) are distributions of the complex amplitudes along the $x$-axis with the split in the $+x$ and $+y$ direction, respectively, where (c,e) for channel $x$ (Ch$x$) and (d,f) for channel $y$ (Ch$y$). The black dashed lines represent the theoretical distributions in the ridge waveguide. RR, CM and CMM represent the rigorous reciprocity, the convolution model and the coupled moment model. CMM2 in (f) stands for CMM that contains the electric quadrupole component.}  
\end{figure*}

For the tip-sample system in Fig. \ref{fig:3}\textcolor{blue}{a}, the detector is right above the SRR, so reciprocal excitation propagates along the $-z$ direction. Besides, excitation in the experimental scenario is beneath the evaluation plane $\Sigma$ so that the noise term $\bm{E}_{\text{rec}}\cdot \bm{I}_{\text{exp}}$ disappears. Here, three models based on reciprocity theory are adopted for analysis and comparisons, including the rigorous reciprocity (RR, cf. Appendix \ref{sec:appendix_c}), the convolution model (CM, cf. Appendix \ref{sec:appendix_c}) and the CMM. We simulate the reciprocal fields of SRR in CM and RR by finite element method (COMSOL Multiphysics 5.2), and calculate the reciprocal dipole moments of SRR by finite-difference time-domain method (FDTD Solutions 8.11) (cf. Fig. \ref{fig:3}\textcolor{blue}{b}). Moreover, the experimental fields in RR are pointwise calculated by moving SRR along $x$-axis with a displacement step of 250 nm (COMSOL Multiphysics 5.2). All the simulation results by RR, CM and CMM are illustrated in Figs. \ref{fig:3}\textcolor{blue}{(c-f)}.

As shown in Fig. \ref{fig:3}\textcolor{blue}{b}, whatever the polarization direction of the reciprocal excitation is, an obvious reciprocal electric dipole moment always exists, indicating that the SRR is sensitive to the in-plane electric fields. Besides, a significant reciprocal magnetic dipole moment $m_{\rm{tip,}\;z}$ appears when the polarization direction becomes perpendicular to the facet of the split (i.e., $E_x$ generates $m_{\rm{tip,}\;z}$), which is the reciprocal process of the Faraday$'$s law and qualitatively interprets the reason for probing $B_z$. Quantitatively, comparing the complex signals $\cal{M}_{\textit{x}}$ in channel $x$ (Ch$x$) with split along $x$ direction (cf. Fig. \ref{fig:3}\textcolor{blue}{c}) with those in Ch$x$ with split along $y$ direction (cf. Fig. \ref{fig:3}\textcolor{blue}{e}), one obtains the dominant $E_x$ distribution along $x$ direction. All the amplitude and phase distributions in RR, CM and CMM are consistent with those of $E_{\textit{x}}$ in the fundamental TE mode of RW, regardless of the constant offsets in distributions of the relative phase. This is a solid demonstration for the effectiveness of the proposed CMM, indicating the validity of the perturbation assumption involved in CM and CMM. 

Strikingly, the comparisons between the complex signals $\cal{M}_{\textit{y}}$ in channel $y$ (Ch$y$) with split along $x$ direction (cf. Figs. \ref{fig:3}\textcolor{blue}{d}) and those in Ch$y$ with split along $y$ direction (cf. Figs. \ref{fig:3}\textcolor{blue}{f}) lead to the conclusion that SRR is sensitive to and can detect the verticle magnetic field $B_z$, which is most obvious in the central part of the amplitude distribution and in the whole distribution of the relative phase. Experimental results in Ref. [\citenum{RN309}] well corroborate the tendency and distributions of these simulations, demonstrating the feasibility and effectiveness of CMM again. Here, we can re-examine the process of optical magnetism detection from the view of power coupling by CMM. In the reciprocal scenario, after the downward $y-$polarized plane wave excites the SRR with split along $x$ direction, one part of power transfers to the reciprocal electric dipole moment $p_{\rm{tip}\;y}$, and the other part generates loop current density in SRR, inducing an effective reciprocal magnetic dipole moment $m_{\rm{tip}\;z}$. In turn, in the experimental scenario, $p_{\rm{tip}\;y}$ and $m_{\rm{tip}\;z}$ couple to $E_y$ and $B_z$ of the fundamental TE mode of RW. However, the $E_y$ component is much weaker than $B_z$ in the central part. Thus, most of the power carried by the far-field $y-$polarized detected light results from the magnetic power coupling. This is the reason why a split-ring probe can detect $B_z$ of the RW.

Finally, we discuss the discrepancy between CMM and RR or CM (cf. left panel of Fig. \ref{fig:3}\textcolor{blue}{f}). Calculation results by CM (in blue triangle) closely resemble those by RR (in red circles) even near the edge of the ridge (i.e., $x=\pm 1 \;\mu \text{m}$) due to the smoothing effect by convolution with adjacent fields. However, the distribution by CMM in Eq. (\ref{equ_8}) (in green squares) is very sharp at the edge of the ridge. This discrepancy may originate from the truncation of  reciprocal multipolar moments. The sharp distribution of electric field at the edge of the ridge could lead to a prominent gradient of electric field $\bm{\nabla E}$. Hence, the distribution is further calculated by CMM that also considers the reciprocal quadrupole moment tensor $\bm{Q}_{\rm{tip}}$ (denoted as CMM2). This new distribution (in purple stars) mostly removes the discrepancy near the edge of the ridge, implying that $\bm{Q}_{\rm{tip}}$ should be included by CMM for higher precision where the sample field has a significant $\bm{\nabla E}$.

\subsection{Transverse SAM detection by a nanoparticle probe}
Except for the commonly discussed longitudinal SAM, light also possesses transverse SAM (tSAM). This unusual tSAM has connection with photonic spin-orbit interaction and contributes to the spin-directional locking phenomenon in nanophotonics \cite{RN32,RN20}. Thus, tSAM is a key vectorial quantity of light to be characterized. Though mostly accompanying the evanescent part of light \cite{RN92,RN70}, tSAM is also discovered in propagating light in two-wave interference \cite{RN268}. Here, the detection of tSAM in tightly focused cylindrical-vector beams by a nanoparticle probe inspired by Refs. [\citenum{RN87,RN303}] is analyzed based on CMM. To describe the tSAM, we start with the general definition of the SAM \cite{RN32,RN87}, which satisfies $\bm{s}\varpropto\text{Im}[\varepsilon_0(\bm{E^\ast}\times\bm{E})+\mu_0(\bm{H^\ast}\times\bm{H})]$. This definition depends on both the electric and magnetic fields, and can be written explicitly as
\begin{equation} \label{equ_10}
\left\{
\begin{aligned}  
s^{\;x}&\varpropto\text{Im}[E^\ast_yE_z+Z_0^2H^\ast_yH_z] \\
s^{\;y}&\varpropto\text{Im}[E^\ast_zE_x+Z_0^2H^\ast_zH_x] \\
s^{\;z}&\varpropto\text{Im}[E^\ast_xE_y+Z_0^2H^\ast_xH_y] 
\end{aligned}
\right. ,
\end{equation}
with superscripts $i=x,y,z$ denoting the spinning axes, where $x$ and $y$ correspond to the tSAM, while $z$ corresponds to the longitudinal SAM.

Two strategies have already been proposed for the detection of tSAM. One is the use of Mie particles as local probes to detect the mechanical action generated by tSAM \cite{RN268}, and the other applies nanoparticles (NPs) as local probes to detect the spin-directional scattering by tSAM with the back focal plane imaging technique \cite{RN87,RN303}. In the latter strategy, the electric tSAM $\bm{s}_{E}$ was successfully detected by gold NP (GNP) \cite{RN87}, and the magnetic tSAM $\bm{s}_H$ was recently measured by silicon NP (SiNP) \cite{RN303}.For detecting or distinguishing the electric or magnetic tSAM, NPs should be either electric- or magnetic-sensitive [cf. Eq. (\ref{equ_10})]. GNP with diameter less than 100 nm can be considered as a solid electric dipole (ED) at visible spectrum, which makes GNP a standout candidate as probes for electric field of light \cite{RN323}. Compared with metallic SRR, NP with high refractive index, especially SiNP, has a strong magnetic dipole (MD) resonance together with low loss \cite{RN96,RN324}. Besides, SiNP with diameter between 100 nm and 200 nm makes the MD resonance cover the whole visible band and can be conveniently fabricated by laser ablation method \cite{RN100}. All these advantages make SiNP widely used in nanophotonics \cite{RN188,RN284,RN120} and a favorable selection as sensors for optical magnetism.

\begin{figure*}
\centering\includegraphics{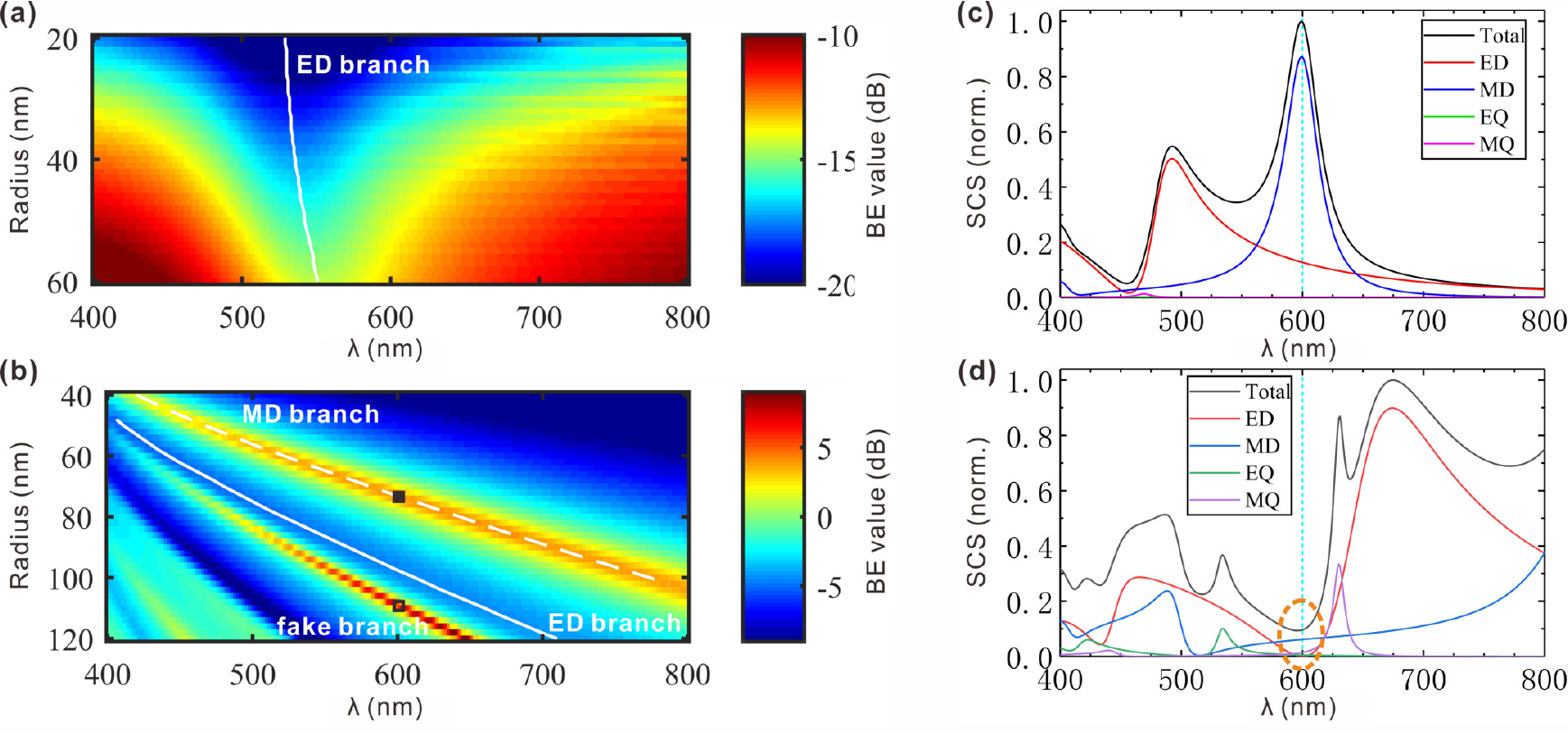}
\caption{\label{fig:4} $BE(\omega)$ of gold and silicon nanoparticles. (a) $BE(\omega)$ of gold nanoparticles (GNPs) with different radii. The solid white line locates electric dipole (ED) resonances. (b) $BE(\omega)$ of silicon nanoparticles (SiNPs) with different radii. The dashed white line and solid white line locate the magnetic dipole (MD) and ED resonances, respectively. (c,d) Scattering cross section (SCS) for different multipole components calculated by Mie theory for SiNP with radius 78 nm and SiNP with radius 112 nm, respectively. The dotted line in (c) indicates location of solid black square in MD branch in (b). The dashed orange circle in (d) shows multipole distributions at the location of open black square in fake branch in (b).}
\end{figure*}

For detection of the tSAM, two criteria ought to be satisfied for the design of nanoprobes. First, the probe should at least have two approximate components of the reciprocal dipole moments according to Eqs. (\ref{equ_10}). NPs are appropriate choices because their geometries are close to nanospheres. Second, for clearly extracting the magnetic tSAM $\bm{s}_H$ from the total tSAM $\bm{s}$, the FOM $BE(\omega)$ is applied to assess the vectorial response of NPs, e.g., GNP and SiNP. Figure \ref{fig:4} illustrates the calculation results of $BE(\omega)$ by FDTD and the theoretical results of resonant wavelengths by Mie theory \cite{RN1169}. For GNP with radius less than 60 nm (cf. Fig. \ref{fig:4}\textcolor{blue}{a}), $BE(\omega)<-  10\;\rm{dB}$, showing that GNP is electric-sensitive at visible spectrum and the resonance wavelengths (i.e., extrema in $BE$) of ED are in accordance with those calculated by Mie theory. For SiNP with radius between 40 nm and 120 nm (cf. Fig. \ref{fig:4}\textcolor{blue}{b}), there are three significant branches in the distributions of $BE(\omega)$ according to the calculations by Mie theory, labeled as the $MD\;branch$, $ED\;branch$ and $fake\;branch$. For MD and ED branches, the maximum and minimum of $BE(\omega)$ correspond to MD and ED resonances of the scattering cross section (SCS) (cf. Fig. \ref{fig:4}\textcolor{blue}{c}), respectively. Thus, SiNP in MD branch is more magnetic-sensitive, satisfying $BE(\omega)\sim 5\;\rm{dB}$, and is an effective sensor for magnetic field of light. On the contrary, SiNP in ED branch is more electric-sensitive, satisfying $BE(\omega)\sim -5\;\text{dB}$, and can be regarded as a sensor for electric field of light. For the fake branch, the situation is a little complicated. Though SiNP in this branch has a higher $BE(\omega)$, satisfying $BE(\omega)\sim10\;\rm{dB}$, it cannot be applied as a magnetic sensor. As depicted in Fig. \ref{fig:4}\textcolor{blue}{d}, the higher $BE(\omega)$ stems from a weaker ED moment rather than a stronger MD moment, where multipole moments [e.g., electric quadrupole (EQ) moment, magnetic quadrupole (MQ) moment, etc.] have comparable strengths with the MD moment. Thus, equation (\ref{equ_9}) is no longer the appropriate truncation of Eq. (\ref{equ_8}) so that $BE(\omega)$ has less physical meaning and makes the corresponding branch named $fake$.

\begin{figure*}
\centering\includegraphics{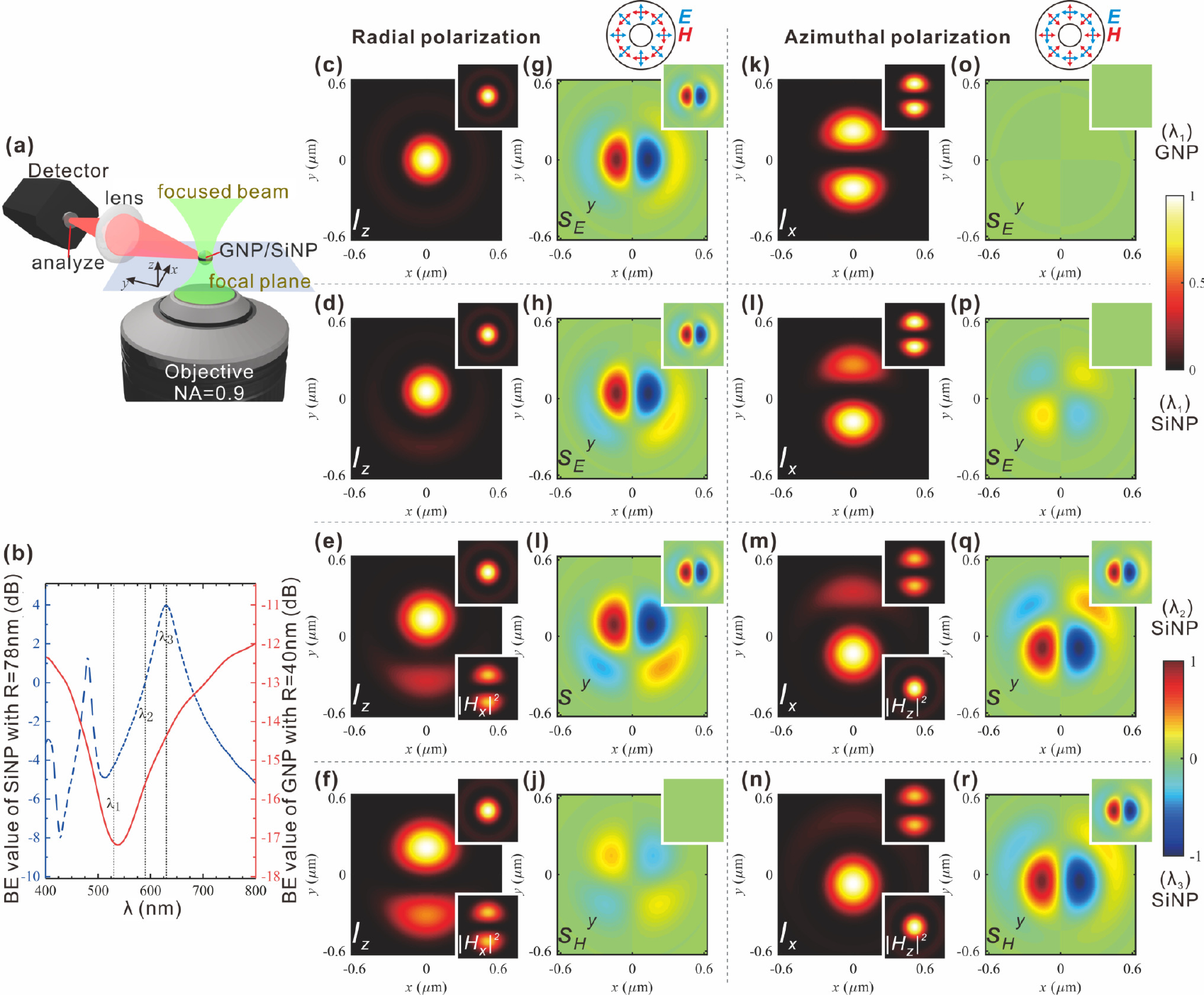}
\caption{\label{fig:5}Simulations of intensity and transverse spin angular momentum (tSAM) detection with nanoparticles (NPs) based on CMM. (a) Optical set-up. NP approaches to the focal plane of the cylindrical-vector beams and scatters light to the far-field detector. (b) $BE(\omega)$ of the selected GNP (solid red line) and SiNP (dashed blue line). Dotted lines in (b) locate the wavelengths using in (c-r) where $\lambda_1$ = 530 nm for Row 1 [(c,g,k,o)] and Row 2 [(d,h,l,p)], $\lambda_2$ = 590 nm for Row 3 [(e,i,m,q)] and $\lambda_3$ = 630 nm for Row 4 [(f,j,n,r)]. (c-f) and (g-j) depict the intensity \textit{I$_z$} and corresponding tSAM for radially polarized light, while (k-n) and (o-r) show the counterparts for azimuthally polarized light. Insets represent the theoretical distributions of intensity and tSAM.}
\end{figure*}

The striking feature of SiNP is that it can be either magnetic- or electric-sensitive by choosing different radii and working wavelengths according to the distribution of $BE(\omega)$ (cf. Fig. \ref{fig:4}\textcolor{blue}{b}). Ideally, extremely high $BE(\omega)$ can be realized by suppressing the multipole and ED moments except for the MD moment, which has been theoretically achieved at near-infrared spectrum by optimally designing Au core-Si shell nanostructures \cite{RN157}. Here, GNP and SiNP are chosen as probes to respond to electric and magnetic fields at visible wavelengths, and simulation results by CMM are compared with the experimental distributions in Refs. [\citenum{RN87,RN303}]. We do not adopt the back focal plane imaging technique in these reference. Rather, we propose a third strategy to probe the tSAM by an apertureless SNOM scheme (cf. Fig. \ref{fig:5}\textcolor{blue}{a}): NP is attached to the tip of an elongated optical fiber so that the movement of NP can be manipulated efficiently. To keep the general aim of this subsection, we restrict our analysis to vectorial responses of NP itself and leave the influence of scattering from the tiny fiber tip to further work. By switching the polarization direction of the analyzer between the $z$ and $x$ axes, signals $E_z \;and/or\; H_x$ and $E_x \;and/or\; H_z$ can be detected by two consecutive SNOM measurements. Then, after postprocessing by Eqs. (\ref{equ_10}), one can obtain tSAM $s_E^{\;y} \;and/or\; s_H^{\;y}$. The above-mentioned strategy can be quantitatively interpreted as follows. In the two reciprocal scenarios, the reciprocal dipole moments approximately satisfy $p_x=p_z=p$ and $m_z=-m_x=m$ according to the spherical symmetry of NPs. Thus, after two orthogonal detections, one obtains $\mathcal{M}_x \approx -i\omega pE_x + i\omega m c^{-1} Z_0 H_z$, $\mathcal{M}_z \approx -i\omega pE_z - i\omega m c^{-1} Z_0 H_x$ [cf. Eq. (\ref{equ_8})] and the corresponding calculated quantity
\begin{equation} \label{equ_12}
\begin{split}
\text{Im}\{\mathcal{M}_z^\ast \mathcal{M}_x\} \varpropto\;&+\text{Im}\{ E_z^\ast E_x \} + 10^{0.2BE}Z_0^2 \text{Im}\{ H_z^\ast H_x \} \\
&+ 10^{0.1BE}Z_0 \text{Im}\{ e^{i\theta}H_x^\ast E_x \} \\
&- 10^{0.1BE}Z_0 \text{Im}\{ e^{-i\theta}E_z^\ast H_z \},
\end{split}
\end{equation}
where \textit{BE} is the BE value in dB for NPs and $\theta$ is the phase difference between the ED moment \textit{p} and the MD moment \textit{m}  [i.e., $\theta = \arg (p) - \arg (m)$]. Particularly, for tightly focused cylindrical-vector beams (Appendix \ref{sec:appendix_d}), $E_z$ and $H_z$ cannot simultaneously exist. Hence, the last term in Eq. (\ref{equ_12}) vanishes, and the exact principle of our strategy to detect tSAM with NPs is
\begin{equation}             \label{equ_13}
\begin{split}
\text{Im}\{\mathcal{M}_z^\ast \mathcal{M}_x\} \varpropto\;&+\text{Im}\{ E_z^\ast E_x \} + 10^{0.2BE}Z_0^2 \text{Im}\{ H_z^\ast H_x \} \\
&+ 10^{0.1BE}Z_0 \text{Im}\{ e^{i\theta}H_x^\ast E_x \}.
\end{split}
\end{equation}

Similar parameters taken from Refs. [\citenum{RN87,RN303}] are used as follows. The radially- (RPL) and azimuthally polarized light (APL) are tightly focused by an objective with numerical aperture 0.9, whose actual fields in focal plane are analytically calculated by MATLAB. The radii of GNP and SiNP are 40 nm and 78 nm, respectively. The reciprocal ED and MD moments are calculated by FDTD excited by plane waves with incident angle $75^{\rm{o}}$ relative to $z$-axis. For fully evaluating the vectorial response of the selected SiNP, three working wavelengths are considered, i.e., $\lambda_1$ = 530 nm, $\lambda_2$ = 590 nm and $\lambda_3$ = 630 nm, while GNP only works at $\lambda_1$ = 530 nm (cf. Fig. \ref{fig:5}\textcolor{blue}{b}).

First, when NP is much more sensitive to the electric field, i.e., \textit{BE} $<$ -5 dB, the coefficient of the first term in Eq. (\ref{equ_13}) is larger than those of the others by one order of magnitude. Thus, according to Eqs. (\ref{equ_10}), one obtains
\begin{equation}    \label{equ_14}
\text{Im}\{\mathcal{M}_z^\ast \mathcal{M}_x\} \varpropto \;\text{Im}\{ E_z^\ast E_x \}\varpropto \;s_E^{\;y}
\end{equation}
with subscripts $E$ or $H$ identifying the prime components of tSAM $s^{\;y}$. SiNP and GNP at $\lambda_1$  satisfy the above-mentioned condition (cf. Fig. \ref{fig:5}\textcolor{blue}{b}) and can be applied to detect the electric tSAM $s_E^{\;y}$. For GNP at $\lambda_1$, the simulated tSAM $\text{Im}\{\mathcal{M}_z^\ast \mathcal{M}_x\}$ by CMM closely resembles the theoretical tSAM $s_E^{\;y}$, both for RPL (cf. Fig. \ref{fig:5}\textcolor{blue}{g}) and APL (cf. Fig. \ref{fig:5}\textcolor{blue}{o}), and obviously coincides with the corresponding experimental results in Ref. [\citenum{RN87}]. For SiNP at $\lambda_1$, the intensity and simulated tSAM distributions [cf. Figs. \ref{fig:5}\textcolor{blue}{(d,h,l,p)}] are in analogy with those of GNP [cf. Figs. \ref{fig:5}\textcolor{blue}{(c,g,k,o)}]. Hence, just like GNP, SiNP working near its ED resonance can be regarded as an effective local probe for electric near fields and corresponding tSAM $\bm{s}_E$.

Second, when NP is much more sensitive to the magnetic field, i.e., \textit{BE} $>$ 5 dB, the second term in Eq. (\ref{equ_13}) becomes dominant. Thus, according to Eqs. (\ref{equ_10}), one obtains
\begin{equation}    \label{equ_15}
\text{Im}\{\mathcal{M}_z^\ast \mathcal{M}_x\} \varpropto \;Z_0^2\text{Im}\{ H_z^\ast H_x \}\varpropto \;s_H^{\;y}.
\end{equation}
SiNP at $\lambda_3$ approaches the above-mentioned condition (cf. Fig. \ref{fig:5}\textcolor{blue}{b}) and can be used to detect the magnetic tSAM $s_H^{\;y}$. In this situation, the intensity and simulated tSAM distributions for RPL [cf. Figs. \ref{fig:5}\textcolor{blue}{(f,j)}] and APL [cf. Figs. \ref{fig:5}\textcolor{blue}{(n,r)}] are complementary to those of GNP or SiNP at $\lambda_1$, implying that SiNP at $\lambda_3$ is magnetic-sensitive. Besides, distributions of the simulated tSAM by CMM for RPL (cf. Fig. \ref{fig:5}\textcolor{blue}{j}) and APL (cf. Fig. \ref{fig:5}\textcolor{blue}{r}) are also consistent with the experimental results in Ref. [\citenum{RN303}], demonstrating the validity and effectiveness of CMM. Hence, SiNP working near its MD resonance can be regarded as a solid nanoprobe for magnetic near fields and the corresponding tSAM $\bm{s}_H$.

Third, when NP has a nearly equal sensitivity to the electric and magnetic fields, i.e., \textit{BE} $\approx$ 0 dB, according to Eqs. (\ref{equ_10}) and (\ref{equ_13}), one obtains
\begin{equation}     \label{equ_16}
\text{Im}\{\mathcal{M}_z^\ast \mathcal{M}_x\} \varpropto \;s^{\;y} + Z_0 \text{Im}\{ e^{i\theta}H_x^\ast E_x \}.
\end{equation}
Equation (\ref{equ_16}) implies the striking potential to measure the complete tSAM $s^{\;y}$ (i.e., both $s_E^{\;y}$ and $s_H^{\;y}$), if and only if the last term in Eq. (\ref{equ_16}) is negligible compared with $s^{\;y}$ in magnitude. As for applications with tightly focused cylindrical-vector beams, the longitudinal fields are much larger than the transverse fields in the central part (i.e., $|H_z|>|H_{x/y}|$ for APL, or $|E_z|>|E_{x/y}|$ for RPL). Therefore, the term $Z_0 \text{Im}\{ e^{i\theta}H_x^\ast E_x \}$ can be regarded as a perturbation term superposing the actual tSAM $s^{\;y}$. SiNP at $\lambda_2$ satisfies the above-mentioned condition of $BE(\omega)$ (cf. Fig. \ref{fig:5}\textcolor{blue}{b}) and can be selected to detect the complete tSAM $s^{ y}$. As illustrated in Figs. \ref{fig:5}\textcolor{blue}{(i,q)}, distributions of simulated tSAM by CMM are in agreement with those of the calculated $s^{\;y}$, both for RPL and APL. Besides, the simulated tSAM for RPL (cf. Fig. \ref{fig:5}\textcolor{blue}{i}) and APL (cf. Fig. \ref{fig:5}\textcolor{blue}{q}) are in analogy with that for RPL (cf. Fig. \ref{fig:5}\textcolor{blue}{h}) and that for APL (cf. Fig. \ref{fig:5}\textcolor{blue}{r}), respectively, except for a minor difference which might result from the last term in Eq. (\ref{equ_16}). Hence, SiNP working at specific wavelength, which satisfies $BE(\omega)\approx0\;\text{dB}$, can be regarded as a local probe for either electric or magnetic near fields of light. More importantly, such SiNP can be an extraordinary candidate for probing of the complete tSAM $\bm{s}$.

\section{Conclusions}
In summary, we have thoroughly investigated the imaging theory for probing the vectorial near fields of light. Our study is based on the rigorous reciprocity of electromagnetism that connects the response of the probe with the actual fields determined by the probe-field interaction. By introducing the multipole expansion analysis to the nanoprobes, the proposed imaging theory can be expressed in the formulation of multipolar series in scalar product. This formulation is in close analogy with the multipole Hamiltonian for the light-matter interaction energy, and interprets the vectorial imaging process from the view of power coupling. This finding is an intrinsic extension to the prior electric dipole model of the probe, evidencing the commonly disregarded magnetic counterpart and revealing possibility of probing the intriguing gradient of electric fields. The proposed imaging theory also provides a quantitative underpinning for analyzing the vectorial response of the probe, addressing the problem whether the dominant contribution of the SNOM signal originates from electric or magnetic fields, and even predicting which component(s) will be picked up by the probe. Specifically, both the vectorial response of the probe and the probe-field interaction would become complicated provided that the complexity of the probe increases. In this work, we restrict our concentration on the electromagnetic response of the probes and leave the analysis of the probe-field interaction to further research. Thus, we developed an approximated model, the coupled moment model (CMM), by applying the perturbation assumption \cite{RN167} to ease the requirement of computational resources.

When designing a nanoprobe for imaging the vectorial near fields, there is a huge parameter space to explore. Therefore, sufficient FOMs should be provided to predict or optimize the probe. The $reciprocal\;dipole\;moments$ of the probe $\bm{p}_{\rm{tip}}$ and $\bm{m}_{\rm{tip}}$ that contain information of both the polarizability of the probe and the SNOM scheme, and the FOM $BE(\omega)$ directly derived from CMM are all solid choices to guide a practical design. According to the above FOMs, a paradigm with established principles was proposed for the designs of functional nanoprobes. In general, the probing of electric near field prefers a probe with larger $\bm{p}_{\rm{tip}}$ and more negative $BE(\omega)$, while a probe with larger $\bm{m}_{\rm{tip}}$ and more positive $BE(\omega)$ suits the detection of magnetic near fields. Moreover, we have numerically demonstrated two applications of vectorial imaging by using the proposed imaging theory and design principles, i.e., the probing of optical magnetism by a split-ring probe \cite{RN309} and the detection of transverse spin angular momentum by a nanoparticle probe \cite{RN87, RN303}. The consistencies of distributions between the calculations and reported experiments \cite{RN309, RN87, RN303} indicate the validity and effectiveness of the proposed theory and principles. It is envisioned that this work could accelerate the development of novel functional probes and novel schemes or techniques in near-field microscopy, and open new avenues for probing the vectorial near fields with new physical quantities in both the theoretical and experimental research in near-field optics and nano-optics.

\begin{acknowledgments}
This work was supported by the National Natural Science Foundation of China (NSFC) under Grant No. 61775113, 11474180, and 61227014. The authors thank Tong Cui for technique support in the numerical simulations.
\end{acknowledgments}

\appendix

\section{\label{sec:appendix_a}Scattering theory and derivation of equation (\ref{equ_2})}
Supposing the background medium is vacuum, one can obtain the vector wave equation issued from Maxwell's equations for monochromatic harmonics:
\begin{equation} \label{equ_26}
-\bm{\nabla}\times\bm{\nabla}\times\bm{E} + k_0^2\bm{E} = -\omega^2\mu_0\varepsilon_0(\bm{\widetilde{\varepsilon}}_r - {\bm{I}}) \cdot\bm{E},
\end{equation}
where $\bm{I}$ is the unit tensor. This expression may be cast as
\begin{equation} \label{equ_27}
-\bm{\nabla}\times\bm{\nabla}\times\bm{E}(\bm{r}) + k_0^2\bm{E}(\bm{r}) = - i\omega\mu_0\bm{J}_\text{P}(\bm{r}),
\end{equation}
where $\bm{J}_\text{P}$ is the polarization current density. The kernel function or Green dyadics $\bm{G}$ is introduced to solve Eq. (\ref{equ_27}), which satisfies
\begin{equation} \label{equ_28}
-\bm{\nabla}\times\bm{\nabla}\times\bm{G}(\bm{r},\bm{r}') + k_0^2\bm{G}(\bm{r},\bm{r}') = - \bm{I}\delta(\bm{r}-\bm{r}')
\end{equation}
and the associated solution:
\begin{equation} \label{equ_29}
\left\{
\begin{aligned}
&\bm{E}_\text{s}(\bm{r}) = i\omega\mu_0\oint_{V'}{d^3r'\bm{G}(\bm{r},\bm{r}')\cdot\bm{J}_\text{P}(\bm{r}')}\\
&\bm{H}_\text{s}(\bm{r}) = \oint_{V'}{d^3r'[\bm{\nabla}\times\bm{G}(\bm{r},\bm{r}')]\cdot\bm{J}_\text{P}(\bm{r}')}.\\
\end{aligned}
\right.
\end{equation}
This implicit solution is self-consistent and in the form of the Lippmann-Schwinger equation. The Green dyadics is generated by a scalar Green function or spherical wave function as
\begin{equation} \label{equ_30}
\begin{split}
\bm{G}(\bm{r},\bm{r}') &= \bigg{(}\bm{I} + \frac{1}{k_0}\bm{\nabla}\bm{\nabla}\bigg{)}G(\bm{r},\bm{r}')  \\
&= \bigg{(}\bm{I} + \frac{1}{k_0}\bm{\nabla}\bm{\nabla}\bigg{)}\frac{\exp(ik_0|\bm{r} - \bm{r}'|)}{4\pi|\bm{r} - \bm{r}'|}.
\end{split}
\end{equation}
The scalar Green function can be expressed in angular spectrum representation, known as the Weyl identity \cite{RN290}
\begin{equation} \label{equ_31}
G(\bm{r},\bm{r}') = \frac{i}{8\pi^2}\int{d^2K}\bigg{\{}\frac{1}{\gamma}\exp[-i\gamma(z - z')+i\bm{K}\cdot(\bm{R} - \bm{R}')]\bigg{\}},
\end{equation}
where $\gamma (\bm{K}) = \sqrt{k_0^2 - |\bm{K}|^2}$ and $z' \geqslant z$ in this paper. This integral is extended to $0<|\bm{K}|<+ \infty$, containing both the propagating $(|\bm{K}| \leqslant k_0)$ and evanescent $(|\bm{K}|>k_0)$ waves.

After little algebra with no approximation, one can obtain
\begin{eqnarray} \label{equ_32}
&&-\bm{\nabla}\cdot(\bm{E}_2\times\bm{H}_1 - \bm{E}_1\times\bm{H}_2) \nonumber\\
=&& \frac{1}{i\omega\mu_0}\big{[}\partial_i(\epsilon_{ijk}E_{2,j}\epsilon_{kmn}\partial_mE_{1,n}) - \partial_i(\epsilon_{ijk}E_{1,j}\epsilon_{kmn}\partial_mE_{2,n})\big{]} \nonumber\\
=&& \frac{1}{i\omega\mu_0}\partial_i(E_{2,j}\partial_iE_{1,j} - E_{1,j}\partial_iE_{2,j}),
\end{eqnarray}
where $\epsilon_{ijk}$ with subscript $\{i,j,k\}=\{x,y,z\}$ is the three-dimensional Levi-Civita symbol. Inserting Eq. (\ref{equ_32}) into the Lorentz reciprocity $- \bm{\nabla}\cdot[\bm{E}_2\times\bm{H}_1 - \bm{E}_1\times\bm{H}_2] = \bm{E}_2\cdot\bm{J}_{1,\text{ext}} - \bm{E}_1\cdot\bm{J}_{2,\text{ext}}$ with external dipole source $\bm{J}_{1(2)} = \bm{I}_{1(2)}\delta(\bm{r}-\bm{r}_{b(a)})$ and integrating it over the volume \textit{V}' above the plane $\Sigma$ (cf. Fig. \ref{fig:1}), with the divergence theorem one obtains
\begin{equation} \label{equ_33}
\begin{split}
\bm{E}_{2}\cdot\bm{I}_{1}|_{\bm{r}_{b}} - \bm{E}_{1}\cdot\bm{I}_{2}|_{\bm{r}_{a}} =& \int_{V'}{d^3\bm{\nabla}\cdot(\bm{E}_2\times\bm{H}_1 - \bm{E}_1\times\bm{H}_2}) \\
=& \int_{\Sigma}dS[\hat{\textbf{e}}_z\cdot(\bm{E}_2\times\bm{H}_1 - \bm{E}_1\times\bm{H}_2)] \\
=& \frac{1}{i\omega\mu_0}\int_{\Sigma}{dS(E_{2,j}\partial_{z}{E_{1,j}} - E_{1,j}\partial_{z}{E_{2,j}})}.
\end{split}
\end{equation}

From now on we use terms $\bm{S}_{E}(\bm{R},z_0)$ and $\bm{T}(\bm{r})$ to replace $\bm{E}_2(\bm{R},z_0)$ and $\bm{E}_1(\bm{r})$, which mean electric field in the evaluation plane $\Sigma$ in the experimental scenario and electric field in the reciprocal scenario, respectively. According to Eq. (\ref{equ_29}), one can cast field $\bm{T}(\bm{r})$ in the form 
\begin{equation} \label{equ_34}
\begin{split}
\bm{T}(\bm{r}) =& k_0^2\oint_{V'}d^3r'\bm{G}(\bm{r},\bm{r}')\cdot\bm{\chi}(\bm{r}')\cdot\bm{T}(\bm{r}') \\
=& k_0^2\oint_{V'}d^3r'\hat{\textbf{e}}_jT_m(\bm{r}')\chi_{mi}(\bm{r}')\bigg{(}\delta_{ij}+\frac{\nabla_i\nabla_j}{k_0^2}\bigg{)}G(\bm{r},\bm{r}').
\end{split}
\end{equation}
Inserting Eq. (\ref{equ_34}) into Eq. (\ref{equ_33}), and transforming $\bm{S}_{E}(\bm{R},z_0)$ and $\bm{T}(\bm{r})$ into their corresponding angular spectrum representation according to Eqs. (\ref{equ_3}) and (\ref{equ_31}), respectively, one derives the following expression after some algebraic manipulations 
\begin{equation} \label{equ_35}
\begin{split}
&\frac{1}{i\omega\mu_0}\oint_{\Sigma}dS(E_{2,j}\partial_zE_{1,j} - E_{1,j}\partial_zE_{2,j}) \\
=& -i\omega\varepsilon_0\int d^3r'\Bigg{\{} T_m(\bm{r}')\chi_{mi}(\bm{r}')\bigg{(}\delta_{ij} + \frac{\partial_i'\partial_j'}{k_0^2}\bigg{)}S_{E,j}^{(+)}(\bm{r}')\Bigg{\}}
\end{split}
\end{equation}
with the Dirac $\bm{\delta}$ function of angular spectra for simplification using relation $\int d^2R\exp\Big{[}i(\bm{K} + \bm{K}')\cdot\bm{R})\Big{]} = 4\pi^2\delta(\bm{K} + \bm{K}')$. In Eq. (\ref{equ_35}), the term $S_{E,j}^{(+)}(\bm{r}')$ is the synthetic field generated by $\bm{S}_{E}(\bm{R},z_0)$ propagating upwards from plane $z=z_0$ to plane $z=z'$ regardless of the probe. Thus, this term satisfies Gauss's law [i.e., $ \partial_j'S_{E,j}^{(+)}(\bm{r}') = 0$], and Eq. (\ref{equ_35}) becomes
\begin{equation} \label{equ_36}
\begin{split}
\frac{1}{i\omega\mu_0}\oint_{\Sigma}&dS(E_{2,j}\partial_zE_{1,j} - E_{1,j}\partial_zE_{2,j}) \\
&= -i\omega\varepsilon_0\int{d^3r'\{T_m(\bm{r}')\chi_{mj}(\bm{r}')S_{E,j}^{(+)}(\bm{r}')\}} \\
&= \int_{V'}d^3r'\bm{S}_{E}^{(+)}(\bm{r}')\cdot\bm{J}_{\textup{tip,P}}(\bm{r}').
\end{split}
\end{equation}
Inserting Eq. (\ref{equ_36}) into Eq. (\ref{equ_33}), one obtains Eq. (\ref{equ_2}).

\section{\label{sec:appendix_b}Derivation of equation (\ref{equ_5})}
Transforming $\bm{S}_E^{(+)}(\bm{r}')$ in Eq. (\ref{equ_36}) into its angular spectrum representation, one obtains without any approximation 
\begin{equation}              \label{equ_44}
\begin{split}
\frac{1}{i\omega\mu_0}\int_{\Sigma}dS(E_{2,j}&\partial_z{E_{1,j}} - E_{1,j}\partial_z{E_{2,j}}) \\
= \frac{1}{2\pi}\int{d^2K}\Big{\{}&\exp(i\bm{K}\cdot\bm{R}_{\textup{tip}})\bm{S}_E^{(+)}(\bm{K},z_{\textup{tip}}) \\
&\cdot\Big{[}\int{d^3r'}\bm{J}_{\textup{tip,P}}(\bm{r}')\exp(i\bm{k}\cdot\bm{r}')\Big{]}\Big{\}}.
\end{split}
\end{equation}
For probes or optical antennas satisfying the small volume condition, i.e., $|\bm{k}||\bm{r}| \leqslant 1$, it is of physical significance to expand the term $\exp (i\bm{k}\cdot\bm{r}')$ into Taylor series, and one obtains
\begin{equation}             \label{equ_45}
\begin{split}
 \int{d^3r'} \bm{J}_{\textup{tip,P}}&(\bm{r}')\exp(i\bm{k}\cdot\bm{r}') \\
=& \int{d^3r'}\bm{J}_{\textup{tip,P}}(\bm{r}') + i\bm{k}\cdot\int{d^3r'\bm{r}'\bm{J}_{\textup{tip,P}}(\bm{r}')}  \\
&+ \sum_{m\ge{2}}\frac{i^m}{m!}\int{d^3r'\bm{J}_{\textup{tip,P}}(\bm{r}')(\bm{k}\cdot\bm{r}')^m}.
\end{split}
\end{equation}

Using the vector identity $\bm{\nabla}\cdot(\bm{ab})=(\bm{\nabla}\cdot\bm{a})\bm{b} + (\bm{a}\cdot\bm{\nabla})\bm{b}$, one can construct a dyadics $\bm{J}(\bm{r}')\bm{r}'$, which satisfies $\bm{\nabla}'\cdot [ \bm{J}(\bm{r}')\bm{r}' ]=[ \bm{\nabla}'\cdot\bm{J}(\bm{r}') ] \bm{r}' + \bm{J}(\bm{r}')$. Applying integration by parts to this relation, one obtains
\begin{equation}            \label{equ_46}
\bm{p}_{\textup{tip}} = -\frac{1}{i\omega}\int_{V'}{d^3r'\bm{J}_{\textup{tip,P}}(\bm{r}')},
\end{equation}
where $\bm{p}_\text{tip}$ is the induced electric dipole moment. For the derivation of Eq. (\ref{equ_46}), we use the charge continuity equation
\begin{equation}           \label{equ_47}
\bm{\nabla}'\cdot\bm{J}_{\textup{tip,P}}(\bm{r}')+\frac{\partial}{\partial{t}}\rho_{\textup{tip,P}}(\bm{r}')=0,
\end{equation}
where $\bm{\rho}_\text{tip}$ is the induced charge density. Moreover, one can construct a dyadics $\bm{r}'\bm{J}(\bm{r}')$ and split it up in its anti-symmetric and symmetric part, i.e., $\bm{r}'\bm{J}(\bm{r}') = (1/2)\big{[}{\bm{r}'\bm{J}(\bm{r}')-\bm{J}(\bm{r}')\bm{r}'}\big{]} + (1/2)\big{[}{\bm{r}'\bm{J}(\bm{r}')+\bm{J}(\bm{r}')\bm{r}'}\big{]}$. For the anti-symmetric part, one obtains the following expression for any wavevector $\bm{k}$
\begin{equation}            \label{equ_48}
\begin{split}
\bm{k}\cdot\int_{V'}&{d^3r'\Big{\{}\frac{1}{2}\big{[}{\bm{r}'\bm{J}_{\textup{tip,P}}(\bm{r}')-\bm{J}_{\textup{tip,P}}(\bm{r}')\bm{r}'}\big{]} \Big{\}}} \\
=& \frac{1}{2}\int_{V'}{d^3r'\{\epsilon_{mns}\epsilon_{mjk}\hat{\bf{e}}_{s}r_j'J_{\textup{tip,P,}\;k}(\bm{r}')a_n\}} \\
=& -\frac{1}{2}\bm{k}\times\int_{V'}{d^3r'\{\bm{r}'\times\bm{J}_{\textup{tip,P}}(\bm{r}')\}} = -\bm{k}\times\bm{m}_{\textup{tip}},
\end{split}
\end{equation}
where $\bm{m}_\text{tip}$ is the induced magnetic dipole moment of the tip. For any component of the dyadic $\bm{r}'\bm{r}'$ (i.e., $r_i'r_j'$), the corresponding dyadics $r_i'r_j'\bm{J}(\bm{r}')$ satisfies
$\bm{\nabla}'\cdot [ r_i'r_j'\bm{J}(\bm{r}') ] = r_j'J_i(\bm{r}') + r_i'J_j(\bm{r}') + r_i'r_j'\bm{\nabla}'\cdot\bm{J}(\bm{r}')$. Then, one can construct a third-order tensor $\bm{r}'\bm{r}'\bm{J}(\bm{r}')$, which satisfies $\bm{\nabla}'\cdot [ \bm{r}'\bm{r}'\bm{J}(\bm{r}') ] =\bm{J}(\bm{r}')\bm{r}' + \bm{r}'\bm{J}(\bm{r}') + \bm{r}'\bm{r}'\bm{\nabla}'\cdot\bm{J}(\bm{r}')$. Thus, the symmetric part is associated with an expression by applying integration by parts  
\begin{equation}           \label{equ_49}
\begin{split}
\int_{V'}&{d^3r'\Big{\{}\frac{1}{2}\big{[}\bm{r}'\bm{J}_{\textup{tip,P}}(\bm{r}')  +\bm{J}_{\textup{tip,P}}(\bm{r}')\bm{r}'\big{]} \Big{\}}} \\
&= -\frac{i\omega}{6} \Big{(} \bm{Q}_{\textup{tip}} + \bm{I}\int_{V'}{d^3r'r'^2\rho(\bm{r}')} \Big{)},
\end{split}
\end{equation}
where $\bm{Q}_\text{tip}$ is the induced electric quadrupole moment tensor of the tip. Inserting Eqs. (\ref{equ_46}), (\ref{equ_48}) and (\ref{equ_49}) into Eq. (\ref{equ_45}) leads to the following equation          
\begin{equation}          \label{equ_50}
\begin{split}
 &\int{d^3r'\bm{J}_{\textup{tip,P}}(\bm{r}')\exp(i\bm{k}\cdot\bm{r}')}  \\
=& -i\omega\bm{p}_{\textup{tip}} - i\bm{k}\times\bm{m}_{\textup{tip}} + \frac{1}{6}\omega\bm{k}\cdot\bm{Q}_{\textup{tip}} \\
&+ \frac{1}{6}\omega\bm{k}\cdot\bm{I}\int_{V'}d^3r'r'^2\rho(\bm{r}') + \sum_{m\geq2}\frac{i^m}{m!}\int d^3r'\bm{J}_{\textup{tip,P}}(\bm{r}')(\bm{k}\cdot\bm{r}')^m.
\end{split}
\end{equation}
Inserting this equation into Eq. (\ref{equ_44}) generates separate terms as follows corresponding to terms in the right-hand side of Eq. (\ref{equ_50}) in turn.

For the first term, inverting the angular spectrum representation of the synthetic field $\bm{S}_E^{(+)}(\bm{K},z_{\text{tip}})$ yields
\begin{equation}            \label{equ_51}
\begin{split}
-i\omega\bm{p}_{\textup{tip}}\cdot\frac{1}{2\pi}\int{d^2K}\big{\{}\bm{S}_E^{(+)}(\bm{K},z_{\textup{tip}})\exp(i\bm{K}\cdot\bm{R}_{\textup{tip}})\big{\}} \\
= -i\omega\bm{p}_{\textup{tip}}\cdot\bm{S}_E^{(+)}(\bm{r}_{\textup{tip}}).
\end{split}
\end{equation}

For the second term, using the vector identity
$(\bm{a}\times\bm{b})\cdot(\bm{c}\times\bm{d}) = (\bm{a}\cdot\bm{c})(\bm{b}\cdot\bm{d}) - (\bm{a}\cdot\bm{d})(\bm{b}\cdot\bm{c})$, one obtains
\begin{eqnarray}          \label{equ_52}
&&-\frac{i}{2\pi} \int{d^2K}\big{\{} \bm{S}_E^{(+)}(\bm{K},z_{\textup{tip}})\cdot(\bm{k}\times\bm{m}_{\textup{tip}})\exp{(i\bm{K}\cdot\bm{R}_{\textup{tip}})} \big{\}} \nonumber\\ 
=&& i\omega\mu_0\bm{m}_{\textup{tip}}\cdot\frac{1}{2\pi}\int{d^2K}\big{\{} \bm{S}_H^{(+)}(\bm{K},z_{\textup{tip}})\exp{(i\bm{K}\cdot\bm{R}_{\textup{tip}})} \big{\}}  \nonumber\\
=&& i\omega\mu_0\bm{m}_{\textup{tip}}\cdot\bm{S}_H^{(+)}(\bm{r}_{\textup{tip}}),
\end{eqnarray}
where $\bm{S}_E^{(+)}$ is replaced with its magnetic counterpart $\bm{S}_H^{(+)}$ by the relation
$\bm{S}_E^{(+)}(\bm{K},z_0) = -{(\omega\varepsilon_0)}^{-1}\bm{k}\times\bm{S}_H^{(+)}(\bm{K},z_0)$.
This relation is Amp\`{e}re's law with Maxwell's addition in angular spectrum representation for $\bm{S}_E^{(+)}$ and  $\bm{S}_H^{(+)}$.

For the third term, by constructing the gradient dyadics $\bm{\nabla}\bm{S}_E^{(+)}$, one obtains
\begin{eqnarray}           \label{equ_53}
&& \frac{\omega}{12\pi} \int d^2K \big{\{} (\bm{k}\cdot\bm{Q}_{\textup{tip}})\cdot\bm{S}_E^{(+)}(\bm{K},z_{\textup{tip}})\exp{(i\bm{K}\cdot\bm{R}_{\textup{tip}})} \big{\}} \nonumber\\
=&& \frac{\omega}{12\pi} \int d^2K \big{\{} \bm{Q}_{\textup{tip}}:\bm{k}\bm{S}_E^{(+)}(\bm{K},z_{\textup{tip}})\exp{(i\bm{K}\cdot\bm{R}_{\textup{tip}})} \big{\}} \nonumber\\
=&& \frac{1}{6}\omega\bm{Q}_{\textup{tip}}:\frac{1}{2\pi} \int d^2K \big{\{} (-i\bm{\nabla})\bm{S}_E^{(+)}(\bm{K},z_{\textup{tip}})\exp{(i\bm{K}\cdot\bm{R}_{\textup{tip}})} \big{\}}  \nonumber\\
=&& -\frac{1}{6}i\omega\bm{Q}_{\textup{tip}}:\bm{\nabla}\bm{S}_E^{(+)}(\bm{r}_{\textup{tip}}).
\end{eqnarray}

For the fourth term, by Gauss's law in angular spectrum representation  [i.e., $ k_jS_{E,j}^{(+)}(\bm{K}) = 0$], one obtains
\begin{equation}         \label{equ_54}
\begin{split}
\frac{1}{6}\omega\int d^2K \big{\{} (\bm{k}\cdot\stackrel{\leftrightarrow}{\bm{I}})\cdot\bm{S}_E^{(+)}(\bm{K},z_{\textup{tip}})\exp{(i\bm{K}\cdot\bm{R}_{\textup{tip}})} \big{\}} \\
= \frac{1}{6}\omega\int d^2K \big{\{} \bm{k}\cdot\bm{S}_E^{(+)}(\bm{K},z_{\textup{tip}})\exp{(i\bm{K}\cdot\bm{R}_{\textup{tip}})} \big{\}}=0.
\end{split}
\end{equation}

Using Eqs. (\ref{equ_51}), (\ref{equ_52}), (\ref{equ_53}) and (\ref{equ_54}), equation (\ref{equ_44}) yields the reciprocity associated with Hamiltonian as  Eq. (\ref{equ_5})       
\begin{equation}       \label{equ_55}
\begin{split}
\frac{1}{i\omega\mu_0}\int_{\Sigma}&dS(E_{2,j}\partial_z{E_{1,j}} - E_{1,j}\partial_z{E_{2,j}}) \\
=& - i\omega\bm{p}_{\textup{tip}}\cdot\bm{S}_{\textit{E}}^{(+)}(\bm{r}_{\textup{tip}}) +i\omega\bm{m}_{\textup{tip}}\cdot\bm{S}_{\textit{B}}^{(+)}(\bm{r}_{\textup{tip}}) \\
&- \frac{1}{6}i\omega\bm{Q}_{\textup{tip}}:\bm{\nabla}\bm{S}_{\textit{E}}^{(+)}(\bm{r}_{\textup{tip}}) + \cdots,
\end{split}
\end{equation}
where $\bm{S}_B^{(+)}(\bm{r}_\text{tip}) = \mu_0 \bm{S}_H^{(+)}(\bm{r}_\text{tip})$.

\section{\label{sec:appendix_c}Rigorous reciprocity and the convolution model}
One can obtain the rigorous reciprocity (RR) by integrating the Lorentz reciprocity over the upper hemispheric space
\begin{equation}  \label{equ_37}
\bm{E}_{2}\cdot\bm{I}_{1}|_{\bm{r}_{b}} - \bm{E}_{1}\cdot\bm{I}_{2}|_{\bm{r}_{a}} = \int_{\Sigma}{[\bm{E}_{\text{2}}\times\bm{H}_{\text{1}} - \bm{E}_{\text{1}}\times\bm{H}_{\text{2}}]\cdot d\bm{S}}, 
\end{equation}
where the normal unit vector of $d\bm{S}$ is $+\hat{\bm{\text{e}}}_z$. According to the analysis in the main text, the left terms can be replaced by the term $\cal{M}_{\text{pol}}$$(\bm{r})$. Thus, equation (\ref{equ_37}) can be expanded as
\begin{equation}  \label{equ_38}
\mathcal{M}_{\textup{pol}}\;= \int_{\Sigma}{[E_{2,x}H_{1,y} - E_{2,y}H_{1,x} - E_{1,x}H_{2,y} + E_{1,y}H_{2,x}]dS}.
\end{equation}

The RR provides a general framework for near-field imaging by SNOM without any approximation. However, it cannot be applied in a straightforward way due to the complexity of the experimental fields $\bm{E}_{\text{2}}$ and $\bm{H}_{\text{2}}$, which are simultaneously determined by the probe and sample. For utilizing reciprocity in a more practical way, the perturbation assumption of the probe is commonly used \cite{RN167} and can be expressed as 
\begin{equation}   \label{equ_39}
\left\{
\begin{aligned}   
\psi_{\exp}(\bm{R},z_0;\bm{R}_{\textup{tip}},z_{\textup{tip}}) &\approx \psi_{\textup{sample}}(\bm{R})|_{z=z_0}\\
\psi'_{\textup{rec}}(\bm{R},z_0;\bm{R}_{\textup{tip}},z_{\textup{tip}}) &= \psi'_{\textup{probe}}(\bm{R} - \bm{R}_{\textup{tip}})|_{z=z_0-z_{\textup{tip}}}\\
& = \psi'^{(i)}_{\textup{probe}}(\bm{R}_{\textup{tip}} - \bm{R})|_{z=z_0-z_{\textup{tip}}},\\
\end{aligned}
\right.
\end{equation}
where the wave function $\psi$ represents fields $\{ E_x, E_y, H_x, H_y \}$, and $\psi '$ stands for fields with different electromagnetic form and different spatial index compared with $\psi$ (i.e., interchanging between $E$ and $H$, $and$ between $x$ and $y$). The superscript $i$ in Eq. (\ref{equ_39}) means a two-dimensional spatial inversion operation, i.e., $\psi^{(i)}(\bm{R}) = \psi (-\bm{R})$. The subscripts $sample$ and $probe$ represent the near-fields solely determined by the sample in the experimental scenario and by the probe in the reciprocal scenario, respectively. Then, equation (\ref{equ_39}) leads to
\begin{equation}      \label{equ_40}
\begin{split}
\int_{\Sigma}&dS\psi_{\exp}(\bm{R},z_0;\bm{R}_{\textup{tip}},z_{\textup{tip}})\psi'_{\textup{rec}}(\bm{R},z_0;\bm{R}_{\textup{tip}},z_{\textup{tip}}) \\
&\approx \int_{\Sigma}dS\psi_{\textup{sample}}(\bm{R})|_{z=z_0} \psi'^{(i)}_{\textup{probe}}(\bm{R}_{\textup{tip}} - \bm{R})|_{z=z_0-z_{\textup{tip}}}.
\end{split}
\end{equation}
Considering the definition of a convolution operation $g(x,y)=\psi_{\textup{sample}}(x,y)\ast\psi'^{(i)}_{\textup{probe}}(x,y)$, one can cast Eq. (\ref{equ_40}) in the form
\begin{equation}    \label{equ_41}
\begin{split}
&\int_{\Sigma}dS\psi_{\textup{sample}}(\bm{R})\psi'^{(i)}_{\textup{probe}}(\bm{R}_{\textup{tip}} - \bm{R}) \\
=& \int_{\Sigma}dxdy\psi_{\textup{sample}}(x,y)\psi'^{(i)}_{\textup{probe}}(x_{\textup{tip}} - x,y_{\textup{tip}} - y) = g(x_{\textup{tip}},y_{\textup{tip}}).
\end{split}
\end{equation}
Thus, with the perturbation approximation given by Eq. (\ref{equ_39}), one can use a convolution operation to express reciprocity as 
\begin{equation}     \label{equ_42}
\begin{split}
\int_{\Sigma}dS&\psi_{\exp}(\bm{R},z_0;\bm{R}_{\textup{tip}},z_{\textup{tip}})\psi'_{\textup{rec}}(\bm{R},z_0;\bm{R}_{\textup{tip}},z_{\textup{tip}}) \\
&\approx \psi_{\textup{sample}}(\bm{R}_{\textup{tip}})|_{z=z_0}\ast\psi'^{(i)}_{\textup{probe}}(\bm{R}_{\textup{tip}})|_{z=z_0-z_{\textup{tip}}}.
\end{split}
\end{equation}
Inverting $\psi$ and $\psi '$ to near-fields  $\{ E_x, E_y, H_x, H_y \}$ leads to the convolution model (CM)
\begin{equation}  \label{equ_43}
\begin{split}
\mathcal{M}_{\textup{pol}} \approx &+ \Big{[}E_{\textup{sample},x}\ast H_{\textup{probe},y}^{(i,-)} + H_{\textup{sample},x}\ast E_{\textup{probe},y}^{(i,-)}\Big{]}  \\
 &- \Big{[}E_{\textup{sample},y}\ast H_{\textup{probe},x}^{(i,-)} + H_{\textup{sample},y}\ast E_{\textup{probe},x}^{(i,-)}\Big{]}, 
\end{split}
\end{equation}
where the superscript $(-)$ means the reciprocal fields. The CM is effectively and widely used in aperture-type SNOM \cite{RN327,RN199,RN153,RN202} and implies that the near-field imaging by probe is a convolution process, which is in close analogy with the coherent imaging theory in Fourier optics \cite{RN343}. Noteworthily, there are totally a set of four point spread functions (PSFs) for the probe in CM of near-field imaging.

\section{\label{sec:appendix_d}Formulas for the focused cylindrical-vector beams}
The analytic formulas for cylindrical vector beams by focusing of a high numerical aperture objective can be expressed as follows \cite{RN337}. For the radially polarized light: 
\begin{equation} \label{equ_56}
\left\{
\begin{aligned}
E_{\rho} &=& A\int_{\alpha_1}^{\alpha_2}\cos^{\frac{1}{2}}\theta{\sin2\theta}{J_1(k\rho{\sin\theta})}e^{ikz\cos\theta}d\theta    \\  
E_z &=& 2iA\int_{\alpha_1}^{\alpha_2}\cos^{\frac{1}{2}}\theta\sin^2\theta{J_0(k\rho\sin\theta)}e^{ikz\cos\theta}d\theta           \\ 
H_{\phi} &=& \frac{2}{Z}A\int_{\alpha_1}^{\alpha_2}\cos^{\frac{1}{2}}\theta\sin\theta{J_1(k\rho\sin\theta)}e^{ikz\cos\theta}d\theta  
\end{aligned}
\right. ,
\end{equation}
and for the azimuthally polarized light: 
\begin{equation} \label{equ_56}
\left\{
\begin{aligned}
E_{\phi} &=& 2A\int_{\alpha_1}^{\alpha_2}\cos^{\frac{1}{2}}\theta\sin\theta{J_1(k\rho\sin\theta)}e^{ikz\cos\theta}d\theta            \\
H_{\rho} &=& \frac{A}{Z}\int_{\alpha_1}^{\alpha_2}\cos^{\frac{1}{2}}\theta{\sin2\theta}{J_1(k\rho{\sin\theta})}e^{ikz\cos\theta}d\theta  \\
H_z &=& \frac{2iA}{Z}\int_{\alpha_1}^{\alpha_2}\cos^{\frac{1}{2}}\theta\sin^2\theta{J_0(k\rho\sin\theta)}e^{ikz\cos\theta}d\theta          
\end{aligned}
\right. ,
\end{equation}
where $Z$, $z$ and $\theta$ represent the impedance of light, distance from the focal plane and polar angle of light after refraction by the objective. For the simulations in this work, the numerical aperture of the objective equals 0.9 (cf. Ref. [\citenum{RN87}]), and we take full use of this aperture. Thus, the starting angle $\alpha_1$ and the end angle $\alpha_2$ equal to $0^{\text{o}}$ and $64.16^{\text{o}}$, respectively.

\nocite{*}

%

\end{document}